\documentclass[twocolumn,noshowpacs,preprintnumbers,amsmath,amssymb]{revtex4}
\usepackage{graphicx}
\usepackage{dcolumn}
\usepackage{bm}

\newcommand{\beq}{\begin{equation}}
\newcommand{\eeq}{\end{equation}}
\def\be {\begin{equation}}
\def\ee {\end{equation}}
\def\bs#1\es{\begin{split}#1\end{split}}
\def\ba#1\ea{\begin{align}#1\end{align}}
\def\baed#1\eaed{\begin{aligned}#1\end{aligned}}
\def\bged#1\eged{\begin{gathered}#1\end{gathered}}
\def\bea{\begin{eqnarray}}
\def\eea{\end{eqnarray}}

\usepackage{xcolor}
\usepackage[colorlinks = true,
            linkcolor = blue,
            urlcolor  = blue,
            citecolor = blue,
            anchorcolor = blue]{hyperref}

\newcommand{\pd}{\partial}

\newcommand{\fr}{\frac}

\newcommand{\om}{\omega}
\newcommand{\we}{\wedge}

\newcommand{\cT}{\mathcal{T}}

\newcommand{\cK}{\mathcal{K}}
\newcommand{\cN}{\mathcal{N}}

\newcommand{\cV}{\mathcal{V}}

\newcommand{\cZ}{{\mathcal{Z}}}

\newcommand{\cref}{{\bf [check ref]}}

\newcommand{\al}{\alpha}

\begin{document}
\preprint{IPMU19-0017}
\title{  {\Large$\alpha'$}-Corrections and de Sitter Vacua - a Mirage\,?
}
\author{ Matthias Weissenbacher}
\affiliation{%
Kavli Institute for the Physics and Mathematics of the Universe, University of Tokyo, Kashiwa-no-ha 5-1-5, 277-8583, Japan
}


\begin{abstract}
In this work we analyze  the role of $\alpha '$-corrections to type IIB orientifold compactifications in K\"ahler moduli stabilization and inflation.
In particular, we propose  a model independent scenario to achieve  non-supersymmetric Minkowski and de Sitter vacua for geometric backgrounds with positive Euler-characteristic and generic number of K\"ahler moduli. The vacua are obtained by a tuning of the flux superpotential.
Moreover, in the one-modulus case we argue for a mechanisms  to achieve  model independent slow-roll.

\end{abstract}
\maketitle

\section{Introduction}

It is of great interest to find de Sitter vacua of supergravities and string theory  from a phenomenological perspective as  a small positive cosmological constant is observed experimentally,  as well as from a fundamental theoretical point of view.
\newline
No-go theorems such  as \cite{Maldacena:2000mw}  suggest that those cannot be obtained from the  lowest order in $g_s$ and $\alpha'$ of the ten and eleven-dimensional  supergravity action, i.e.\,\,in string or M-theory.  Where $g_s$ is the string coupling constant and  $2 \pi \alpha'= l_s^2$ with $l_s$  being the string length. The recent swampland conjecture \cite{Obied:2018sgi} may be evaded along those lines. It is thus of great interest to advance beyond the classical results \cite{Cicoli:2018kdo}.
In this work we analyze  the role of the leading order $\alpha '$-corrections to the four-dimensional, $\cN =1$  effective actions of type IIB orientifold compactifications \cite{Becker:2002nn,Bonetti:2016dqh,Grimm:2013gma,Grimm:2013bha,Weissenbacher:2019,Ciupke:2015msa,Grimm:2017okk} in K\"ahler moduli stabilization \cite{Westphal:2006tn,Weissenbacher:2019,Ciupke:2015msa}. 
We argue that due to the presence of the $\alpha '$-corrections Minkowski and de Sitter vacua may be obtained for Calabi-Yau orientifolds of positive Euler-characteristic. The proposed scenario assumes that the complex structure moduli are stabilized beforehand  by the flux superpotential \cite{Gukov:1999ya,Sethi:2017phn,Denef:2008wq} and additionally requires the vacuum value of the flux superpotential to be tuned to a specific value in relation to other topological quantities of the Calabi-Yau orientifold \cite{Rummel:2011cd}. Whilst maintaining moderately large values a fine tuning of the flux superpotential engineers  a small cosmological  constant. Let us stress that the vacua are  obtained model independently. However, we do not study explicit geometries in this work, but derive additional constraints on certain topological quantities of the geometric background. Let us emphasize that the vacua generically are at large volume where higher-order $\alpha'$ and $g_s$-corrections are under control and moreover non-perturbative effects to the superpotential are suppressed.

Secondly, we propose a mechanism in which the  $\alpha'$-scalar potential exhibits slow-roll inflation, where the K\"ahler modulus plays the role of the inflaton \cite{Broy:2015zba,Cicoli:2016chb}. We discuss the one-modulus case explicitly and derive model independent results.
To stabilize the overall volume in a Minkowski or de Sitter Minimum at the end of inflation we require the flux superpotential to be perturbed from its vacuum expectation value during inflation. Inflation ends as the flux superpotential acquires its vacuum expectation value and the K\"ahler modulus is stabilized in the  resulting minimum discussed in the first part of this work. 
For simplicity we  do not discuss a dynamical model of this process.  Two possible alternatives occur to us. Firstly, the fluxes may tunnel from one vacuum configuration to another one close by. Secondly, the fluxes are fixed and the complex structure moduli are perturbed slightly from their minimum.
We conclude that one may neglect those  dynamical effects  if  the complex structure modulus perturbed is stabilized at a large complex structure point. Thus the single field approximation is valid and  one may infer the  observable quantities such as the spectral index $n_s$ and the tensor to scalar ratio $r$. For simplicity we work under the assumption  that the flux superpotential perturbation is  constant during inflation and then acquires its vacuum value rapidly.
We focus on the regime $n \approx 0.965$  \cite{Aghanim:2018eyx,Komatsu:2010fb}. For sixty e-foldings we find the tensor-to-scalar ratio to be $r \leq  5.65 \cdot 10^{-4}$. Relaxing this constraint  to  fifty e-foldings one  acquires  $r  \leq  1.19 \cdot 10^{-3}$.

Let us shortly outline the structure of this note. In section \ref{sec-scalarpot} we review the relevant leading order $\alpha'$-corrections to four-dimensional $\cN=1$ super gravity obtained from type IIB orientifolds. We also provide a critical assessment on their status.
In section \ref{sec-oneMod} we argue for the Minkowski and de Sitter vacua in the one-modulus case, followed by section \ref{sec-Generic} in which we show that Minkowski  and de Sitter vacua can be obtained for geometric backgrounds with an arbitrary number of K\"ahler moduli. Lastly, in section \ref{sec-toyInflation} we  propose an inflationary scenario in the one-modulus case.

\section{ {\large$\alpha'$}--corrected scalar potential}\label{sec-scalarpot}

Let us set the stage by reviewing the recent progress made in determining $\alpha'$-corrections to the four-dimensional $\cN=1$ scalar potential of IIB effective actions of string theory.  At the time being three different sources are worth noting. 
The well known Euler-characteristic $\al '^3$-correction to the K\"ahler potential
\beq\label{BBHL}
 \xi =  -  \frac{\zeta(3)}{4 }\,g_s^{{-\tfrac{3}{2}}} \, \chi(Y_3) \;\; , 
 \eeq
with $\chi(Y_3)$ Euler-characteristic of the Calabi-Yau background $Y_3$. Note that it is of order $ \mathcal{O}(\alpha^{\prime3})$ and it depends on 
the Type IIB string coupling,  $g_s = \langle e^{\phi_d }\rangle$ where $\phi_d$ is the dilaton.  It is obtained from the parent $\cN=2$ theory arising  from compactification of type IIB on Calabi--Yau orientifolds \cite{Becker:2002nn,Bonetti:2016dqh}.
Secondly, evidence has been provided for the existence of a $\alpha'^2$-correction to the K\"ahler coordinates utilizing F-theory based on a series of papers \cite{Grimm:2013gma,Grimm:2013bha,Grimm:2014efa,Grimm:2015mua,Grimm:2017pid,Weissenbacher:2019}.  This formulation of Type IIB string theory with  space-time filling  seven-branes at varying string coupling \cite{Vafa:1996xn} captures the string coupling dependence in the geometry of an elliptically fibered Calabi--Yau fourfold with base $B_3$.  Effective actions of F-theory compactifications have been studied using the duality with M-theory \cite{Denef:2008wq,Grimm:2010ks}.  Note that  the type IIB Calabi-Yau threefold is  a double cover of the base $B_3$ branched along the $O7$-plane. Thus  in particular one  infers that $\chi(Y_3) = \chi(B_3)$. The  conjectured $\alpha'^2$-correction to the K\"ahler coordinates \cite{Weissenbacher:2019} is found to be
\beq\label{Z_corr}
T_i = \cK_i +  \;\; \gamma \, \cZ_{i} \text{log}(\cV)\;\; , \;\; \;\;  \text{with}\;\;\; \cZ_i = \int_{D_i } \mathcal{C}  \;\; ,
\eeq
with $T_i$ the K\"ahler coordinates, i.e.\,\,the complex structure on K\"ahler moduli space, and where $i=1,\dots,h^{1,1}(B_3)$ and $\gamma =- \tfrac{1}{64}$.  Note that other $\alpha'^2$-corrections  to the K\"ahler potential and coordinates do not lead to a breaking of the no-scale condition  \cite{Cicoli:2007xp,Weissenbacher:2019}. Thus \eqref{Z_corr} is sufficient for the context of this work. Moreover, we denote the K\"ahler moduli fields as $v^i$ and $\mathcal{C}$ is the curve corresponding to the self-intersection of stacks of $D7$-branes and $O7$-planes \cite{Grimm:2013bha}. Furthermore, $D_i \subset B_3$ denotes divisors in the internal space.  Note that the orientifold involution in IIB Calabi--Yau threefolds with additional  $O7$ and $O3$-planes acting on $Y_3$ projects the K\"ahler moduli space to $H^{1,1}_{+}(Y_3)$, i.e.\,the $(1,1)$-forms that are even under the isometric involution \cite{Grimm:2010ks}, i.e.\, $h^{1,1}(B_3) = h_+^{1,1}(Y_3)$. Let us speculate on the type IIB origin of a logarithmic correction in the volume such as \eqref{Z_corr}.  When four-dimensional minimal supergravity is integrated out at one-loop on the circle the  three-dimensional K\"ahler coordinates receive  an analogues logarithmic correction in the radius \cite{Tong:2014era}. One might expect that \eqref{Z_corr} arises  by integrating out certain  massive Kaluza-Klein states in  a supergravity theory on Calabi--Yau orientifolds.

 We summarize the other relevant topological quantities such as intersection numbers $\cK_{ijk}, \, \cK_{ij},\, \cK_{i} $ and the volume $\cV$ in appendix \ref{Computation}.   Note that  we have defined $\cV$ to be dimensionless in units of $(2 \pi \,\alpha')^3$, i.e.\,\,we measure units of length in units of $\sqrt{2 \pi \alpha'}$.
The contribution to the  F-term scalar potential of a four-dimensional $\cN=1$ super gravity theory is
\beq\label{4dFTerm}
V_F = e^{K}\Big( K_{i j} D^i W  \overline{D^j W} - 3 \big| W \big|^2  \Big) \;\; , \;\;
\eeq
with the superpotential $W$, the K\"ahler potential $K (\text{Re} T_i)$ and $K_{ij}  = (\pd_{\text{Re} T_i } \pd_{\text{Re} T_j } K)^{-1}$. Moreover, with the K\"ahler covariant derivative given by
\beq\label{KaerhlerCoord}
D^iW =  \frac{\pd K}{\pd \text{Re} T_i}   W + \frac{\pd W }{\pd \text{Re} T_i}   \;\; ,\;\;\; 
\eeq
 The K\"ahler potential then results in
\beq\label{corr_KaehlerPot}
K \, = \, - 2 \, \text{log}\Big(\cV \,  + \, \hat \xi \Big) \, +  \, \text{log}\Big( \frac{g_s}{2}\Big).
\eeq
In this work we consider the  $h^{2,1}$ complex-structure moduli  as well as the dilaton to be stabilized by the flux superpotential \cite{Gukov:1999ya} which in the vacuum then takes the  constant value $|W_0|> 0$. Thus supersymmetry is broken in the vacuum. A critical assessment of this two step procedure  is provided in e.g.\,\,\cite{Choi:2004sx}.

Lastly, one encounters contributions to the four-dimensional scalar potential  by considering  four-derivative,  $\cN=1$ supergravity \cite{Ciupke:2015msa} resulting in
\beq
 \delta V_F \, = \, - {\large e}^{2 K} T_{ i  j k l}\overline{D^i W}\,\overline{D^j W}\, {D}^k {W}\,{D}^l {W}\;\;.
\eeq
with the contribution to the IIB effective four-dimensional action to be
\ba\label{Gam-corr}
\delta V_F & \; = \; \frac{3 \, |W_0|^4 \, g_s}{4 \cV^4} \, \hat\gamma_2 \, \Gamma_i \, v^i \;\;\; , \;\; \nonumber \\
  \Gamma_i &  = \int_{D_i} c_2  \; \;, \;\;\;\;
\hat\gamma_2 =  \tfrac{11}{576 \, \cdot 4 \pi} \, g_s^{-\tfrac{1}{2}}  \;\; ,
\ea
where $c_2$ is the second Chern-form of the Calabi--Yau orientifold  which may be written as the square of a covariant object, thus $  \Gamma_i  >0$. The precise form of  \eqref{Gam-corr} was inferred by dimensional reduction  of   ten-dimensional IIB $R^4$ -terms \cite{Gross:1986iv} on Calabi--Yau orientifolds in \cite{Grimm:2017okk}.

The scalar potential arising from the $\alpha'$-corrections \eqref{BBHL}, \eqref{Z_corr}, \eqref{corr_KaehlerPot} and \eqref{Gam-corr} in the large volume limit then takes the form
 \beq\label{scalarPot}
V_F = \frac{3 |W_0|^2 g_s}{4 \cV^3}\Big(  3\, \hat\xi + \gamma \cZ_i \, v^i  \, +  \hat\gamma_2 \frac{ \,  |W_0|^2 }{\cV}  \Gamma_i v^i   \Big)\;\; ,
\eeq
where we have defined $\hat \xi = \xi \, /\,3$. We refer to the large volume limit to the regime at large volumes $\cV$ and weak string coupling  such that higher-order $\alpha'$ and $g_s$-corrections as well as non-perturbative instanton effects can be neglected.  Let us emphasize that the correction \eqref{Z_corr} is intrinsically $\cN=1$ in contrast to \eqref{BBHL} and \eqref{Gam-corr}.

Let us comment on the stability of the following scenarios in section \ref{sec-oneMod}  and \ref{sec-Generic}  in the light of \cite{Dine:1985he}. The classical correction to the scalar potential vanishes due to the no-scale condition and thus the leading order $g_s$ and $\al'$-correction determine the vacuum. Higher-order $\alpha'$-corrections are parametrically under control as one stabilizes the internal space at large volumes. Moreover the string coupling constant $g_s$ may be achieved to be parametrically small thus  higher-order string loop corrections can  generically  be neglected. However, as the perturbative corrections depend on the topological quantities of the geometric background higher-order $g_s$-corrections may become relevant in certain vacua.
 Non-perturbative effects to the superpotential due to gaugino condensation or Euclidean $D3$-brane instanton effects play a sub-leading role as those are exponentially suppressed at large volumes. 
Let us close this section with  critical remarks on the status of the $\alpha'$-corrections. A derivation of the well established $\alpha'^3$-correction \eqref{BBHL} in a full-fledged $\cN=1$ set-up such as F-theory is absent \cite{Minasian:2015bxa}. Furthermore, the   $\alpha'^2$-correction \eqref{Z_corr} has been  conjectured recently in \cite{Weissenbacher:2019}, in particular the factor $\gamma$ might be subject to change and thus its  ultimate faith is yet to be determined in future work.
Concerning, the correction \eqref{Gam-corr} two caveats are to be named. Firstly, the four-derivative $\cN=1$ on-shell supergravity action remains elusive \cite{Ciupke:2015msa}. Moreover as pointed out in \cite{Grimm:2017okk} relevant eight-derivative terms in ten-dimensions are not established to fix $\hat\gamma_2$ precisely. It would be of great interest in particular in the light of this work to  investigate  those topics.

\section{One K\"ahler modulus vacua} \label{sec-oneMod}

\begin{figure}[!ht]
     \centering
\begin{center}
\includegraphics[height=4.9cm]{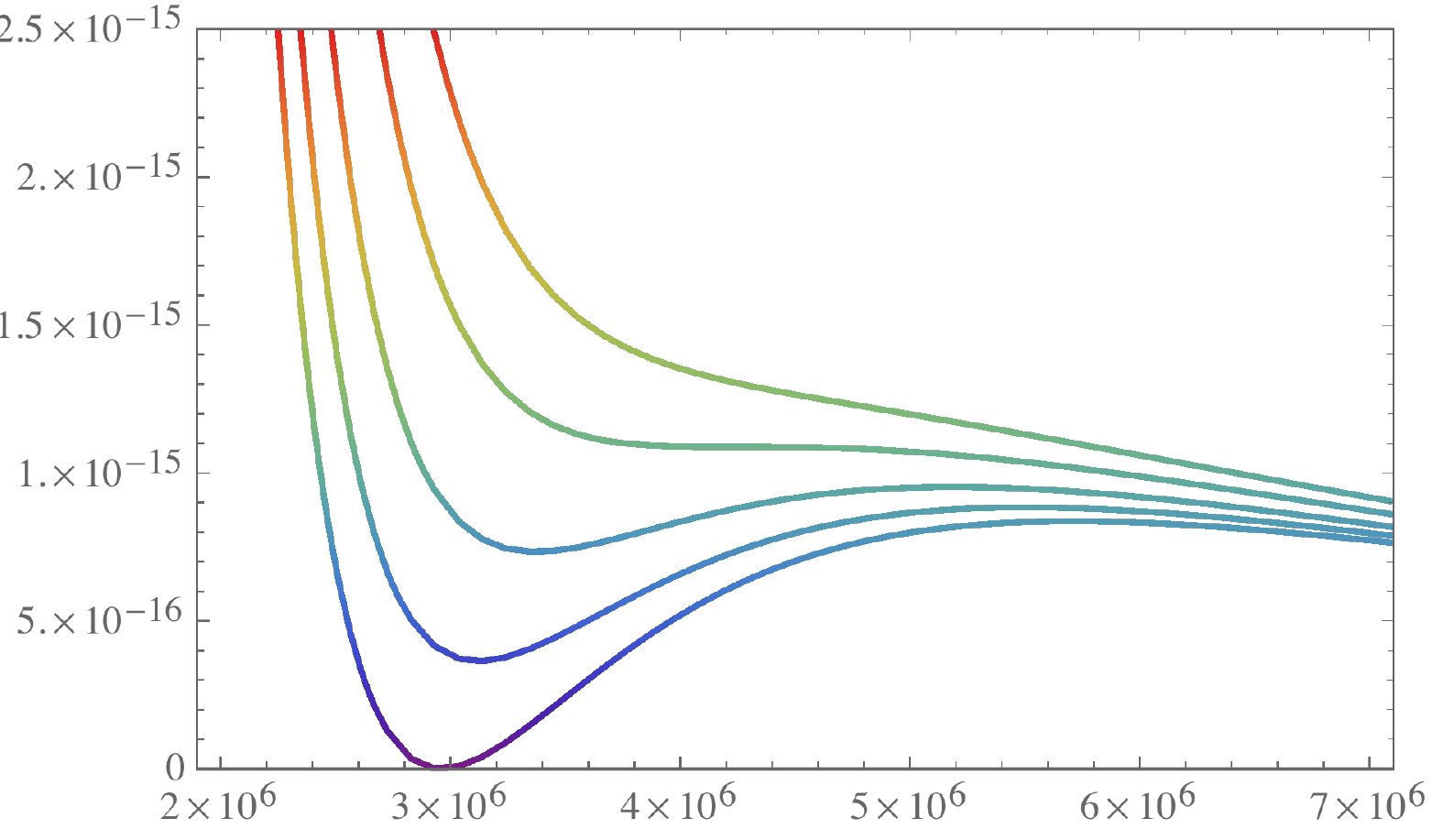}
\end{center}\vspace{-0.3cm}
     \caption{Minkowski and de Sitter vacua. The scalar potential \eqref{oneMod} for the  choice of the string coupling $g_s = 0.2$ and the topological number $\chi=\Gamma = 1$ and $\cZ =-1$. From below with $\Omega = 0, \,  0.5,  \,  0.75,  \,  1 ,  \,  1.2.$. The horizontal  axis displays  the volume $\cV \, /  \, (2 \pi \alpha' )^3$  versus the  scalar potential $V_F$ on the vertical axis. }\label{fig:PlotVacua}
   \end{figure}

In this section we study the one-modulus case  to lay the foundation for the generic scenario discussed in section \ref{sec-Generic}. We show that  the overall volume may be stabilized in a non-supersymmetric de Sitter and Minkowski minima  by the scalar potential \eqref{scalarPot} for manifolds with $\chi(Y_3)> 0$. Note that as $\chi(Y_3) = 2 h^{1,1} - 2 h^{2,1}$ the geometric background  needs to obey $ h^{1,1} > h^{2,1}$ i.e.\,$h^{1,1} > 1$.  However, note that the orientifold involution projects the K\"ahler moduli space to $H^{1,1}_{+}(Y_3)$.  Thus we consider backgrounds with $h^{1,1}_{+} = 1$  in this section. In the one-modulus case the scalar potential  \eqref{scalarPot} derives to
\beq\label{oneMod}
V_F = \frac{3 |W_0|^2 \, g_s}{4\, \cV^3}\Big(   3\hat\xi + \gamma \cZ  \, \cV^{\tfrac{1}{3}} +  \hat\gamma_2 \,  |W_0|^2\,  \Gamma \, \cV^{-\tfrac{2}{3}}  \Big)\;\; .
\eeq
 We argue for  model independent  Minkowski  and de Sitter vacua  which may me achieved by tuning  the  fluxes. The superpotential  in the vacuum  $|W_0|$  is to be fixed at specific values in relation to the topological  $\cZ$, $\chi$ and $\Gamma$. Furthermore, one requires $\cZ < 0$. The vacuum is obtained for
\ba   \label{Mimina}\vspace{-0,2cm}
   \langle \cV \rangle |_{ \Omega = 0}= \frac{8 \, | \hat\xi |^3 }{|\gamma \, \cZ|^3 }  \;\; , \;\;\;    \langle \cV \rangle |_{ \Omega = 1}= \frac{11.3906 \, | \hat\xi |^3 }{|\gamma  \, \cZ|^3 } \;\; ,\\[0.2cm] \label{W0condition}
|W_0|^2 = \frac{|\hat\xi|^3 }{\gamma^2 \, \cZ^2 \, \Gamma \, \hat\gamma_2}\left( 4 + \tfrac{25}{176} \, \Omega^2 \right) \;\;,\;\;\; 0 \leq \Omega^2 \leq 1 \;\; .
\ea 
where $\Omega $ can be chosen freely. In the case of $\Omega = 0$ one finds a Minkowski minimum in \eqref{Mimina}, else-wise de Sitter minima. The  local maximum of the potential is located at
\beq\label{MaximumPot}
\cV_{max}  |_{ \Omega = 0}= \frac{15.3024 \, | \hat\xi |^3 }{|\gamma \,\cZ|^3 } \;\; .
\eeq
 In figure \ref{fig:PlotVacua} we illustrate the scalar potential exhibiting the discussed features. 
Note that  as $ \langle \cV \rangle \,  \, \sim \, g_s^{-{9}/ {2} }$ large volumes can be achieved easily at weak string coupling. Furthermore, the numerical value of $\gamma$  in \eqref{Z_corr} favours large volumes for moderate values of the topological quantities. Let us stress that the minimum is obtained by tuning the fluxes such  that \eqref{W0condition} is satisfied.  The value of the potential in the minimum is
\beq\label{Vac_Min_OneMod}
 \langle V_F \rangle \, = \, f(\Omega)  \cdot \frac{| \gamma \, \cZ|^7}{ \hat\gamma_2 \, \Gamma \, |\hat \xi|^5} \, \sim \, g_s^{9} \;\; , 
\eeq
where on can solve for $f(\Omega)$ analytically.  The explicit expressions for $f(\Omega)$ is long and non-illustrative thus we just note that $f(0) =0$  and $f(1) = 0.000573274$. For weak string coupling small values of the potential in the minimum are natural as one concludes from \eqref{Vac_Min_OneMod}. Let us next compare the gravitino mass with  the string  and Kaluza-Klein scale \cite{Conlon:2005ki} for which one finds that
\beq\label{scalingMasses}
m_S \sim \langle \cV \rangle^{-\tfrac{1}{2}} \;\;\;\;\ ,\;\;\;\;\; m_{KK}  \sim \langle \cV \rangle^{-\tfrac{2}{3}}  \;\;\;\;\ ,\;\;\;\;\; m_{3/2} \sim \frac{|W_ 0|}{\langle\cV\rangle} \;\; .
\eeq
One infers from \eqref{scalingMasses} for the present scenario the relations 
\beq\label{gravitinoOnemoduls}
\frac{ m_{3/2}}{ m_S} = \sqrt{\frac{\gamma \, \cZ}{ 2\hat\gamma_2 \,\Gamma}} \sim  g_s^{\tfrac{1}{4}}\;\; ,\;\;\;\frac{m_{KK}}{m_{S}} \sim g_s^{ \tfrac{3}{4}} \;\; ,
\eeq
which follows from the scaling $|W_0| \, \sim \, g_s^{-2}$ as one infers from \eqref{W0condition}. To achieve a gravitino mass smaller than the Kaluza Klein-scale we note that $ m_{3/2} \, /\,  m_{KK} \,  =  \, (|\hat\xi | \, / \, \hat\gamma_2 \Gamma )^{1/2} \, < 1$, thus the topological quantities need to be accordingly as it scales with $\sim g_s^{-1/2}$.  

Let us close this section with  a short remark on the stability of the found vacua.  The Minkowski vacuum constitutes a global minimum and is thus  stable whilst the de-Sitter vacua are meta-stable. One can estimate the life time of the de Sitter vacua\, i.e.\, the inverse probability for decay in the runaway direction  $\cV \to \infty$, for small $\Omega$  to be
\beq\label{lifetime}
\tau \, \sim \,  \tfrac{1}{\Omega^2} \cdot \text{Exp}\left({ \frac{\gamma^2 \cZ^2}{2 |\hat\xi |\, \sqrt{\hat\gamma_2 \Gamma} \, \Omega^{3/11} } }\right) \, \sim \tfrac{1}{\Omega^2} \cdot e^{\, g_s^{7/4} \, /\,  \Omega^{3/11}  }\;\;.
\eeq 
From \eqref{lifetime} one infers that large life times require a fine tuning of $\Omega \ll 1$ and thus of the superpotential\, i.e.\,the fluxes in \eqref{W0condition}. 
Let us   emphasize that the proposed moduli stabilization scenario  does not require non-perturbative effects but is achieved solely by the leading order $\alpha'$-effects.

\section{Generic K\"ahler moduli scenario}\label{sec-Generic}

In this section we propose  scenarios in which all K\"ahler moduli might be stabilized in a non-supersymmetric de Sitter and Minkowski minima for manifolds with $\chi(Y_3)> 0$. We argue for a model independent extremum and  provide  sufficient conditions  on the topological quantities of the geometric background for the existence of a local minimum. Furthermore, the presented scenarios require a fine tuning of  fluxes such that the value of the superpotential  in the vacuum  $|W_0|$ exhibits a certain relation w.r.t.\,the topological quantities  $\cZ_i$ and $\chi$.
 The stabilization is achieved by an interplay of the $\cZ_i$-correction, with the  $\alpha'^3$ Euler-characteristic correction \eqref{BBHL} to the K\"ahler potential \cite{Becker:2002nn,Bonetti:2016dqh} and the correction to the scalar potential from the four derivative-terms \cite{Ciupke:2015msa}, resulting in the scalar potential \eqref{scalarPot}.
 
 We first discuss the case of Minkowski vacua and then extend the discussion to obtain de Sitter vacua. 
 Let us emphasize that the we do not require non-perturbative effects which are generically exponentially suppressed by the volume of the cycles.
 \newline

 \paragraph{Minkowski vacua. } We henceforth  consider the scalar  potential \eqref{scalarPot}. One may choose K\"ahler cone variables $v^i > 0$ such that the K\"ahler form is
 \beq\label{Kahlerform}
 J = v^i \omega_i \;\; .
 \eeq
 One obtains  Minkowski  vacua for  $\chi(Y_3)> 0$ i.e.\,\,$\xi  < 0$ where all four-cycle volumes $\cK_i$ are stabilized at 
\ba\label{vacuum_one}
\langle \cK_i \rangle &=   \, \frac{ |W_0|^2 \, \hat\gamma_2}{| \hat\xi |} \, \big( \gamma  \,  \Lambda^2 \, \cZ_i \,  + \, \, \Gamma_i \big) \;\; , \\ \label{vacuum_two}
\Lambda^2 &= \frac{1}{|\hat \xi|}  \Gamma_i \langle v^i \rangle \;\; , \;\;\; \cZ_i \langle v^i \rangle = \frac{2 |\hat \xi|}{\gamma}  \;\;\;.
\ea
As $\Gamma_i > 0$ a sufficient condition for positivity of the four-cycle volumes is that $\cZ_i < 0$ for all $i=1,\dots, h_+^{1,1}(Y_3)$.
Note that \eqref{vacuum_one} and \eqref{vacuum_two} constitute an implicit definition for $\langle v^i \rangle$ which generically does not admit a solution. However, we assert that by tuning $|W_0|$  one may always find a solution. To argue in favor of this claim  note that \eqref{vacuum_one} only depends on \eqref{vacuum_two} via a scaling.
By using $ \langle v^i_0 \rangle =\tfrac{1}{| W_ 0|}  \langle v^i \rangle $ one infers that  
\beq\label{defV0}
  \cK_{ijk} \langle v^j_0 \rangle  \langle v^k_0 \rangle  =  \, \frac{\hat\gamma_2}{|\hat \xi |}  \Big(2  \tfrac{  \Gamma_j \langle v^j_0 \rangle}{ \cZ_k \langle v^k_0 \rangle}  \, \cZ_i \,  +   \, \Gamma_i \Big) \;\; ,\;\;\;
\eeq
where one imposes conditions on the fluxes such that
\beq\label{W0condiMinkowski}
|W_0|  =  \frac{ 2|\hat \xi| }{  |\gamma|} \cdot \frac{1}{ |\cZ_i| \langle v^i_0 \rangle}  \;\;.
 \eeq
   Note that  \eqref{defV0} is dimensionless as on the r.h.s.\,the powers of $\alpha'$ cancel. In other words the additional conditions \eqref{vacuum_two} can be satisfied by tuning $|W_0|$  according to \eqref{W0condiMinkowski} if  \eqref{defV0}  exhibits a solution for  $ \langle v^i_0 \rangle $. One infers the volume in the extremum to be
 \beq\label{volumeVac}
\langle \cV \rangle=  \, \frac{|W_0|^2 \, \hat\gamma_2}{|\hat \xi|}   \cdot\Gamma_i \langle v^i \rangle  \,  \sim  \, g_s^{-\tfrac{9}{2}}  \;\;, 
 \eeq
where we have used that $|W_0| \sim g_s^{-2}$.  For moderate values of the topological quantities  \eqref{volumeVac} is thus  generically large. Moreover, we find that the value of the potential vanishes $\langle V^F \rangle =  0$.   Note that as we find a Minkowski minimum we do not need tor require  $|W_0| $  to be small which bypasses certain criticism \cite{Kallosh:2004yh}.  One may easily verify that \eqref{vacuum_one} and \eqref{vacuum_two} constitutes an extremum of the scalar potential  \eqref{scalarPot}.
By analyzing the matrix of second derivatives in the extremum one infers
\ba\label{variation2}
 \Big<  \frac{ \pd^2\, V_F}{\pd v^i \pd v^j}   \Big>  \;\; &=  \frac{3|W_0|^2 |\hat \xi| g_s}{2\, \langle \cV \rangle^4}\Big( \langle \cK_i  \rangle \langle \cK_j \rangle - \langle \cV \rangle \langle \cK_{ij} \rangle + \cT_{ij}\Big)\, ,  \\
\text{with} \; \;\;\;\; & \cT_{ij} = \tfrac{\hat\gamma_2^2}{\hat\xi^2} \Big(  \gamma^2 \Lambda^4 \, \cZ_{i}  \cZ_{j} \, - \,\Gamma_{i}  \Gamma_{j}  \Big) \;\;\;.\label{variation3}
\ea
The objective is to argue that \eqref{variation2} is positive definite. Note that the  positive definite metric on the K\"ahler moduli space  is proportional to $ \langle\cK_i \rangle \langle \cK_j  \rangle-  \langle\cV \rangle \langle \cK_{ij} \rangle $, thus one remains to show the criterium for $\cT_{ij}$. For any vector within  the K\"ahler cone $v^i$ we find the sufficient condition that
\beq\label{conditionVac1}
\cT_{ij} v^i v^j \,  >\,  0 \;\;\;\; \text{if} \;\;\;\; |\cZ_{i}| \, > \,  \tfrac{1}{ |\gamma| \Lambda^2}\Gamma_i \ \;\; ,
\eeq 
for all $ i=1,\dots,h^{1,1}_+(Y_3)$. Thus in geometric backgrounds in which  \eqref{conditionVac1} is  satisfied  one encounters a local Minkowski minimum for all K\"ahler moduli.  One finds
 \beq
 \Lambda^2 = \frac{2}{|\gamma|} \frac{\Gamma_i \langle v_ 0^i\rangle}{|\cZ_i| \langle v_ 0^i\rangle} \,\;\; ,
 \eeq
 and thus the geometric condition \eqref{conditionVac1} becomes
 \beq\label{geometrCond2}
 |\cZ_{i}| \, > \, \frac{1}{2}\frac{|\cZ_i| \langle v_ 0^i\rangle}{\Gamma_i \langle v_ 0^i\rangle}  \Gamma_i \;\; .
 \eeq
 A strong sufficient requirement for \eqref{geometrCond2} to be satisfied is that $\alpha_{max} < 2 \cdot \alpha_{min}$ where we have used that $\Gamma_i = \alpha(i) |\cZ_i|$, for a positive vector with components $\alpha (i) \in \mathbb{Q}_+$. 
 Thus we conclude  that the requirements on the geometric background are that $\chi(Y_3 )> 0$, $\cZ_i \, < \, 0$ for all 4-cycles, and that the inequality \eqref{ds-conditionVac1} is valid. Moreover, one requires the existence of fluxes which can be fine-tuned as \eqref{W0condiMinkowski}.
Thus one infers by using  \eqref{scalingMasses}  and \eqref{volumeVac}   that
\beq
\frac{ m_{3/2}}{ m_S} \sim  \sqrt{\frac{|\gamma|}{2\hat\gamma_2}  \cdot \frac{|\cZ_i| \langle v^i \rangle}{\Gamma_i \langle v^i \rangle}} \sim g_s^{\tfrac{1}{4}}\;\; ,\;\;\;\; \frac{m_{KK} }{ m_{S}} \sim g_s^{ \tfrac{3}{4}}  \;\;.
\eeq
The gravitino mass   is to be smaller than the KK-scale as the effective supergravity theory is derived in a classical reduction. One infers that
\beq\label{garvtinoKK}
m_{3/2} \, < \, m_{KK} \;\;\; \text{if} \;\;\;   \Gamma_i \langle v_ 0^i\rangle \,  > \,  \frac{| \hat \xi |}{ \hat\gamma_2} \;\; .
\eeq
Let us conclude that this mechanism  may lead to a stabilization of all four-cycles  such that the  overall volume is stabilized at sufficiently large value  thus higher-order $\alpha'$-corrections are under control i.e.\,\,the vacuum may  not be shifted.
\newline

 \paragraph{de Sitter vacua. } 
We henceforth consider a modification of the previous  Minkowski vacuum solution such that it constitutes a de Sitter minimum of the  scalar potential \eqref{scalarPot}. We  choose K\"ahler cone variables $v^i > 0$ such that the K\"ahler form  \eqref{Kahlerform} is positive.
We next argue that one obtains  de Sitter  vacua for  $\chi(Y_3)> 0$. i.e.\,$\xi  < 0$ where all four-cycle volumes $\cK_i$ are stabilized at 
\ba\label{ds-vacuum_one}
\langle \cK_i \rangle &=   \, \frac{ |W_0|^2 \, \hat\gamma_2}{|\hat \xi | (1 + \Omega^2 / 8)} \, \left( \frac{ \gamma\,   \Lambda^2}{1 - \Omega^2} \, \cZ_i \,  + \, \, \Gamma_i \right) \;\; , \\ \label{ds-vacuum_two}
\Lambda^2 &= \frac{1}{|\hat \xi|}  \Gamma_i \langle v^i \rangle \;\; , \;\;\; \cZ_i \langle v^i \rangle = \frac{ |\hat \xi|}{\gamma}\left(2 + \tfrac{11}{8} \Omega^2\right)  \;\;\;.
\ea
where we have introduced the positive parameter
\beq
0 \,\, < \,\, \Omega^2 \,\, < \tfrac{2}{11} \;\;.
\eeq
As $\Gamma_i > 0$, a sufficient condition for the positivity of the four-cycle volumes is that $\cZ_i < 0$ for all $i=1,\dots, h_+^{1,1}(Y_3)$.
Note that \eqref{ds-vacuum_one} and \eqref{ds-vacuum_two} analogue to \eqref{vacuum_one} constitute an implicit definition for $\langle v^i \rangle$.  We  next assert that by tuning $|W_0|$  one may always find a solution. Note that \eqref{ds-vacuum_one} only depends on \eqref{ds-vacuum_two} via a scaling.
By using $ \langle v^i_0 \rangle =\tfrac{1}{| W_ 0|}  \langle v^i \rangle $ one infers that  
\beq\label{ds-defV0}
  \cK_{ijk} \langle v^j_0 \rangle  \langle v^k_0 \rangle  =  \, \frac{\hat\gamma_2}{| \hat\xi |(1 + \Omega^2 / 8)}  \Big(2 \gamma_3 \tfrac{  \Gamma_j \langle v^j_0 \rangle}{ \cZ_k \langle v^k_0 \rangle}  \, \cZ_i \,  +   \, \Gamma_i \Big) \;\; ,\;\;\;
\eeq
with $ \gamma_3 = (1 + \tfrac{11}{16}\Omega^2) / (1 - \Omega^2)$ where one imposes
 \beq\label{W0ds}
 |W_0| = \frac{ 2 (1 + \tfrac{11}{16} \Omega^2 ) |\hat \xi|}{ | \gamma | \, |\cZ_i| \langle v^i_0 \rangle } \;\;.
  \eeq
In particular, one infers that  \eqref{ds-vacuum_two} can be satisfied by tuning  the flux-superpotential according to \eqref{W0ds}  if  \eqref{ds-defV0}  exhibits a solution for  $ \langle v^\al_0 \rangle $.   Let us next discuss relevant quantities in the de Sitter extremum. We infer
\beq\label{deSitter_Val}
\langle V_ F \rangle \,\, = \,\, \frac{9\, \Omega^2  }{8} \cdot \frac{|\hat \xi |\, |W_ 0|^2}{ \langle \cV \rangle^3} \, \sim g_s^9  \, > 0 \;\; ,
\eeq
with the volume in the vacuum 
 \beq\label{ds-volumeVac}
\langle \cV \rangle=  \, \frac{|W_0|^2 \, \gamma_2}{| \xi| \big(1 - \Omega^2 \big)}   \cdot\Gamma_i \langle v^i \rangle  \sim g_s^{-\tfrac{9}{2}} \;\;, 
 \eeq
where we have used that $|W_0| \sim g_s^{-2}$ and $\langle v_0^i\rangle \sim g_s^{1/2}$. One infers  from \eqref{deSitter_Val} that in the present scenario a small cosmological constant i.e.\,\,vacuum value $ \langle V_ F \rangle \ll 1$ is achieved naturally due the weakly coupled string regime $g_s < 1$. We expect however that the vacuum stability analysis  of section \ref{sec-oneMod} carries over to the generic moduli case  and thus  a  fine-tuning of the fluxes such that $\Omega^2 \approx 0$ is to be required. 
Let us mention that anti-de Sitter vacua are obtained  for  $\Omega^2 < 0$.
 Furthermore, the positive overall  volume   may generically be  stabilized at large values as seen from \eqref{ds-volumeVac}.
 We are next to show that the extremum \eqref{ds-vacuum_one} and \eqref{ds-vacuum_two} at hand is a local minimum. 
One computes the  matrix of second derivatives in the extremum to be
\ba\label{ds-variation2}
 & \Big<  \frac{ \pd^2\, V_F}{\pd v^i \pd v^j}   \Big>  \;\; =  \frac{3|W_0|^2 |\hat \xi| \, g_s}{2\langle \cV \rangle^4}\Big( G_{ij} + \cT_{ij}\Big)\;\;\;\; , \;\; \\[0.2cm] \label{ds-variation2a}
\;\; & G_{ij} \, =\, \big(1 - \tfrac{35}{8} \Omega^2 \big) \langle\cK_i \rangle  \langle\cK_j \rangle-  \big(1 + \tfrac{1}{8} \Omega^2 \big)\langle \cV \rangle \langle\cK_{ij} \rangle  \;\;,\\
 \;\; &  \cT_{ij}  = \tfrac{\hat\gamma_2^2}{ \hat\xi^2(1 + \Omega^2 /8)}  \Big(  \tfrac{\gamma^2 \Lambda^4}{ (1 - \Omega^2)^2} \, \cZ_{i}  \cZ_{j} \, - \,\Gamma_{i}  \Gamma_{j}  \Big) \;\;\;.\label{ds-variation3}
\ea
The objective is to show that \eqref{variation2} is positive definite.
 It was argued  in \cite{CANDELAS1991455} that $ \langle\cK_{ij} \rangle$ is of signature $(1,h^{1,1}(Y_3))$, i.e.\,it exhibits one positive eigenvalue  in the direction of the vector $\langle v^i \rangle$. As the pre-factor of  $\langle \cK_{ij} \rangle$ is negative and moreover $\langle\cK_i\rangle\langle \cK_j \rangle$ is positive semi-definite, we need to show that  $(G_{ij} + \mathcal{T}_{ij}) \, \langle v^i \rangle \,\langle v^j \rangle > 0$ which leads to $\Omega^2< 2 \, /\, 11$. 
 Thus one remains to show the criterion for $\cT_{ij}$ in generic direction. For any vector within  the K\"ahler cone $v^i$ we find the sufficient condition that 
\beq\label{ds-conditionVac1}
\cT_{ij} v^i v^j \,  >\,  0 \;\;\;\; \text{if} \;\;\;\; |\cZ_{i}| \, > \, \frac{1-\Omega^2}{2 + \tfrac{11}{8} \Omega^2}\frac{|\cZ_i| \langle v_ 0^i\rangle}{\Gamma_i \langle v_ 0^i\rangle}\Gamma_i \;\;,
\eeq 
for all $ i=1,\dots,h_+^{1,1}(Y_3)$. 
One infers that a strong sufficient condition for \eqref{geometrCond2} to be satisfied is that $\alpha_{max} <  (2+11\, \Omega^2\,/\,8) \,/ (1- \Omega^2) \cdot \alpha_{min} $, where we have used that $\Gamma_i = \alpha(i) |\cZ_i|$, for a positive vector with components $\alpha(i)  \in \mathbb{Q}_+$. 
It remains to show that the matrix \eqref{ds-variation2} is positive definite on the entire space, i.e.\,\,for all vectors $\langle v^i \rangle + v^i_\perp $, where $ \langle \cK_i \rangle v^i_\perp=0$ describes the orthogonality. Note that positivity of \eqref{ds-variation3} is granted due to \eqref{ds-conditionVac1}. Thus the non-vanishing contribution which could alter the bound on $\Omega^2$  however computes to a positive number  $-(1 +\Omega^2\, /8)) \langle \cV \rangle \langle \cK_{ij} \rangle  v^i_\perp  v^j_\perp > 0$,  as  $\langle \cK_{ij} \rangle$ admits negative eigenvalues  w.r.t.\,\,$v^j_\perp$ .

Moreover, one concludes assuming \eqref{ds-conditionVac1} that \eqref{ds-variation2} admits one negative eigenvalue, i.e.\,\,not stabilized direction and $h^{1,1}_+(Y_3) -1$ positive eigenvalues for 
\beq\label{boundnflation}
 \tfrac{2}{11} < \Omega^2 < \tfrac{8}{35}\;\; ,
\eeq
which will be of relevance in section \ref{sec-toyInflation}.
 Thus we conclude  that the requirements on the geometric background are that $\chi(Y_3) > 0$,    $\cZ_i \, < \, 0$ for all 4-cycles, and that the inequality \eqref{ds-conditionVac1} is valid, and moreover the existence of fluxes which can be fine-tuned such that \eqref{W0ds} is satisfied.

By using  \eqref{scalingMasses}   one may compare the gravitino mass with  the string  and Kaluza-Klein scale.  One infers that
\beq\label{HirachyScales}
\frac{ m_{3/2}}{ m_S} \sim  g_s^{\tfrac{1}{4}}\;\; ,\;\;\; \frac{ m_{3/2}}{ m_{KK}} \sim   g_s^{ -\tfrac{1}{2}} \;\; ,\;\;\; \frac{m_{KK}}{m_{S}} \sim g_s^{3/4}  \;\;.
\eeq
The gravitino mass achieved needs to be smaller than the Kaluza Klein-scale. A concise analysis gives an analogous relation to \eqref{garvtinoKK} to be
\beq\label{garvtinoKK2}
m_{3/2} \, < \, m_{KK} \;\;\; \text{if} \;\;\;  \Gamma_i \langle v_ 0^i\rangle \,  > \,   \frac{|\hat\xi |}{ \,\hat\gamma_2} \cdot (1 -\Omega^2) \;\; .
\eeq
Note that a small cosmological constant implies that $\Omega^2 \approx 0$ and thus   \eqref{garvtinoKK2}  reduces to   \eqref{garvtinoKK}. 

Let us conclude that  all four-cycles are  generically stabilized such that the  overall volume is at large values if  $\chi(Y_3) > 0$ and $\cZ_i < 0$ for all $i =1,\dots,h_+^{1,1}(Y_3)$.  Thus higher-order $\alpha'$-corrections are under control. Moreover, small string coupling $g_s < 1$ assures correct hierarchies of scales \eqref{HirachyScales} thus the vacuum is to be trusted. Additionally  the constraints on the topological invariants \eqref{ds-conditionVac1} and \eqref{garvtinoKK2} are to be suitable, i.e.\,this imposes  additional restrictions on allowed geometric backgrounds.

\section{A Toy model for  inflation} \label{sec-toyInflation}

In this section we  propose a scenario for inflation based on the one-modulus scalar potential discussed in section \ref{sec-oneMod}. We noted that a minimum of the potential is obtained by tuning the flux in  the vacuum  such that \eqref{W0condition} is satisfied for a choice of $0 \leq \Omega \leq 1$. However the resulting potential does not admit inflationary dynamics.  Intriguingly,  for values $ \Omega > 1$ the potential exhibits features of slow-roll. To achieve a stable final vacuum state we propose  the following scenario. During inflation the flux superpotential is perturbed from it vacuum state such that $ \Omega > 1$. When inflation ends the flux-superpotential acquires its vacuum expectation value such that the inflaton is  stabilized in the Minkowski or de Sitter minimum, i.e.\,\,$\Omega = 0$  and $\Omega \approx 0$, respectively.  In other words we will assume that after a sufficient number of e-foldings, i.e.\,fifty to sixty the flux superpotential obtains its vacuum expectation value which determinates the  inflation process. In, particular in the model at hand the epsilon parameter stays small at high e-folds  especially for $r < 10^{-4} $, thus we need to assume the above mechanism.
Let us note that as we do not study the fluxes dynamically we refer to the described mechanism as scenario.

\begin{figure}[!ht]
     \centering
\begin{center}
\includegraphics[height=6.2cm]{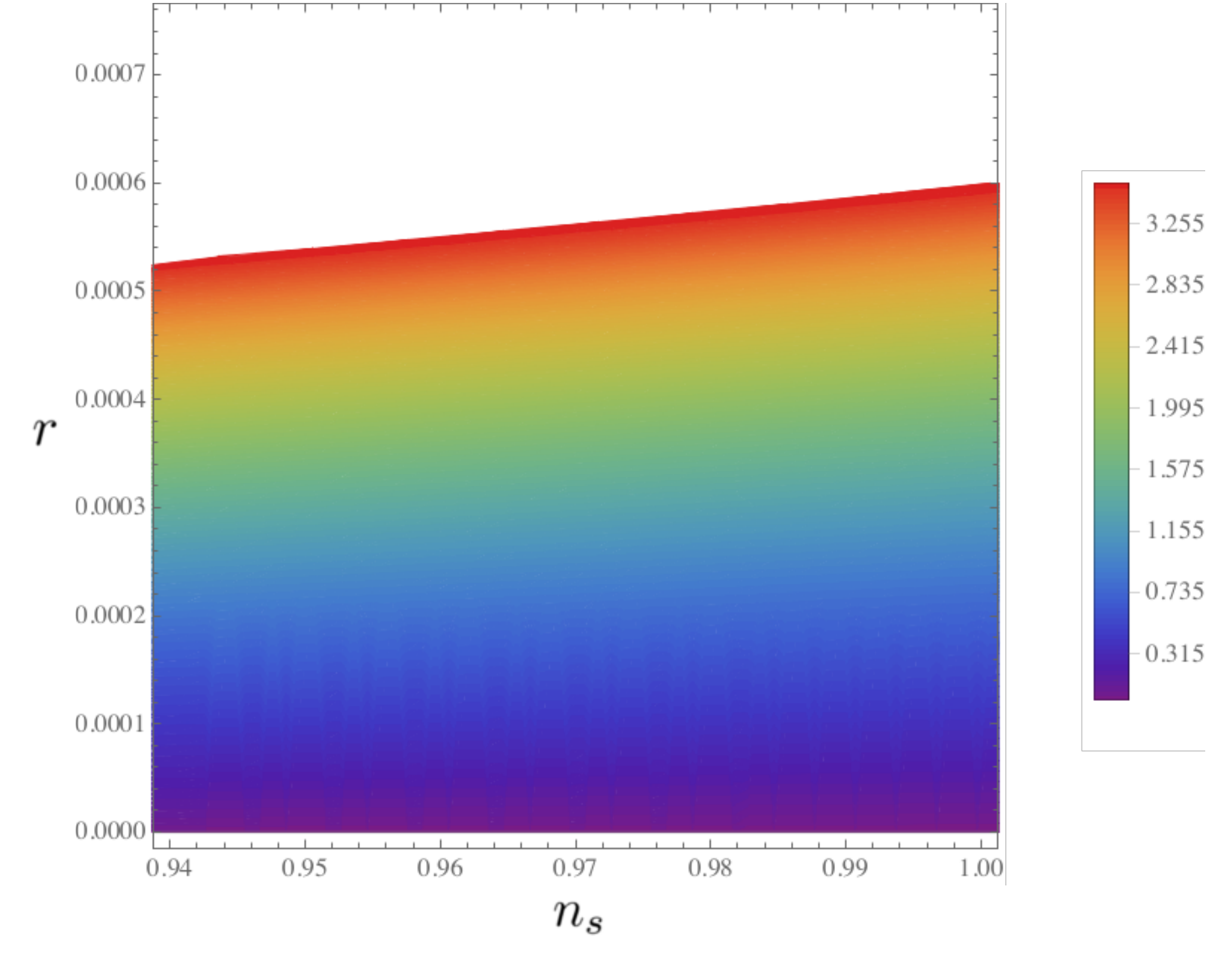}
\end{center}\vspace{-0.3cm}
     \caption{Spectral index  $n_s$ versus tensor-to-scalar ratio $r$ for $N_e = 60$ where we vary the potential parameter in between $0 <\delta \leq 6.5 \cdot 10^{-3}$  to access the entire range shown in the plot. The color theme shows the distance in field space  i.e.\,$ \Delta \phi =\phi_{end} - \phi_\ast$. In the colored region inflation can be accessed. Requiring  less e-foldings e.g.\, $N_e = 50$ one obtains  an analogous result but for increased tensor-to-scalar ratio, bound from above at around  $r \leq 1.19 \cdot 10^{-3}$ at  $n_s \approx 0.965$, for $0 < \delta \leq 9.5 \cdot 10^{-3} $.}\label{fig:PlotNSR}
   \end{figure}

The one K\"ahler modulus with  canonical normalized kinetic term is 
\beq
 \cV = \text{Exp}\Big( \tfrac{\sqrt{3}}{2} \, \Phi \Big) \;\;\; ,\;\;\; \Phi=  \tfrac{2}{\sqrt{3}} \text{Log}(\cV) \;\;,
 \eeq
 where we have suppressed the $\alpha'$-corrections to the kinetic coupling as those lead to sub-leading effects  $\sim 1\, / \cV$.
The inflationary potential \eqref{oneMod} is  thus expressed as
\ba\label{PotINflation1}
V_F &=   \frac{3\, g_s \,|W_0|^2}{ 2\, \text{Exp}\big( \tfrac{11}{\sqrt12}\, \Phi \big)} \left( 3  \hat\xi  e^{\tfrac{\Phi}{\sqrt{3}}}+ \gamma \cZ  \,e^{\tfrac{ \sqrt{3}}{2} \Phi}  +  \hat\gamma_2 \,  |W_0|^2\,  \Gamma  \right) \, , \nonumber\\
&|W_0|^2 = \frac{|\hat\xi|^3 }{\gamma^2 \, \cZ^2 \, \Gamma \, \hat\gamma_2}\left( 4 + \tfrac{25}{176} \,( \Omega^2+  \delta) \right) \;\;,\;\;\; 0 < \delta \ll 1 \; .
\ea
where $\Omega^2 =1$ during inflation. Note that the inflationary potential is to good approximation displayed as the $\Omega=1$ curve in figure  \ref{fig:PlotVacua}.
Let us next comment on the slow-roll parameters
\beq\label{defSlowRoll}
\epsilon \, = \, \frac{1}{2}  \left( \frac{V'_F}{V_F} \right)^2 \;\;\; , \;\;\;\; \eta  \, = \, \frac{V''_F}{V_F} \;\; , 
\eeq
with $V_F' = \pd_\Phi V_F $, and $ V_F''=\pd_\Phi \pd_\Phi V_F$ the partial derivatives w.r.t\,the inflaton. To discuss the slow roll parameters it is convenient to choose coordinates which reflect the minimum and maximum of the potential in the vacuum i.e.\,\,$\Omega \approx 0$ given by \eqref{Mimina} and \eqref{MaximumPot}.  In such a variable $\phi = (|\gamma \cZ| /|\hat\xi|)^3 \cdot \text{Exp}(\sqrt{3} \Phi / 2)$  the slow roll parameters \eqref{defSlowRoll} of the potential \eqref{PotINflation1} become
\ba \label{ep-par}
\epsilon &= \, \frac{1}{24} \left( \frac{ 11\, \hat\delta -27 \, \phi^{2\, / \,3} + 8 \, \phi }{ \hat\delta -3 \, \phi^{2\, / \,3} +  \phi} \right)^2  \;\; ,\\[0.2cm]\label{eta-par}
\eta &= \,  \frac{16}{3} +\frac{ 19\, \hat\delta -17 \, \phi^{2\, / \,3} }{ 4 \, ( \hat\delta -3 \, \phi^{2\, / \,3 } +  \phi) }  \;\; ,
\ea
where $\hat\delta = 4 + \tfrac{25}{176} \,( 1+  \delta) $. Note that for $\delta= 0$  \eqref{ep-par} vanishes at $\phi = 11. 3906$.  We encounter  for $\delta > 0$ that the point of  horizon exit   $\phi_\ast$ is in this region. Furthermore, note that \eqref{ep-par} and \eqref{eta-par} do not depend on the topological quantities of the geometric background and are thus model independent.
The main cosmological observables of interest  are the spectral index and the tensor-to-scalar ratio  evaluated at horizon exit  denoted by the asterisk which are given by
\beq
n_s \, = \, 1 + 2\, \eta_\ast - 6 \, \epsilon_\ast \;\;\; \text{and} \;\;\; r \, = \,  16 \, \epsilon_\ast \;\; ,
\eeq
respectively, i.e.\,$\epsilon_\ast = \epsilon(\phi_\ast)$.  The number of e-foldings in between $\phi_\ast$ and the end of inflation is
\beq
N_e \, = \, \int^{\phi_{end}}_{\phi_\ast} \, \frac{d\, \phi}{\sqrt{2 \, \epsilon (\phi)}}\;\;,
\eeq
as we find $\phi_{end} > \phi_\ast$. The end of inflation  is marked when the $\epsilon$-parameter fails to be much smaller than one \cite{PhysRevD.29.2162}. As we do not discuss a dynamic model for the fluxes acquiring their vacuum value, we simply assert that \, $\delta \to 0$ and $\Omega^2 \to 0$ for Minkowski or for de Sitter some small value $\Omega^2 \ll 1$. We assume this process to be instantaneous to determinate inflation such that $N_e = 60$ is obtained, see figure \ref{fig:PlotNSR}. 
   Note that the field displacement is sufficiently small such that $\phi_{end}$ is well below the maximum of the barrier which forms at $\phi \approx 15.3 $ as seen from \eqref{MaximumPot}. This constitutes a necessary requirement for stabilization of  the modulus in the minimum.
Furthermore, the  density perturbation  amplitude at the  horizon exit is to matched with the observed value to give
\beq
\mathcal{A}_{COBE} (\phi_\ast) \, = \,\   \left. \frac{V^3_F}{(V'_F)^2} \right|_{\phi_\ast}  \,  = \, 2.7 \cdot 10^{-3} \;\; ,
\eeq
which evaluated for the potential \eqref{scalarPot} results in
\beq
    \frac{ 54 \, g_s\, \gamma \cZ \,  |W_0|^2 \, \cV_\ast^{-\tfrac{11}{3 \sqrt{3}}} \Big(  \cV_\ast^{ \tfrac{ 1} {\sqrt{3}}}  -  \tfrac{3\, |\hat\xi| }{ \gamma \cZ} \,  \cV_\ast^{ \tfrac{2} {3\sqrt{3}}} + \cV_{\hat\delta}  \Big)^3}{2.7 \cdot 10^{-3}  \, \Big( 8\, \cV_\ast^{ \tfrac{ 1} {\sqrt{3}}}  - \tfrac{27\, | \hat\xi| }{ \gamma \cZ} \,  \cV_\ast^{ \tfrac{2} {3\sqrt{3}}} +11 \cV_{\hat\delta}   \Big)^2}      =  1 \, , \label{Acobe}
\eeq
 where $\cV_\ast$ is the volume at the horizon exit point and $\cV_{\hat\delta} =   \hat \delta \cdot |\hat\xi|^3\, / \, \gamma^3 \cZ^3 $.
Note that \eqref{Acobe} may be satisfied by stabilizing the dilaton in the vacuum accordingly i.e.\, by the string coupling constant. For the exemplary values $\cZ = -3$ and $ \chi = \Gamma = 1$ one  infers from \eqref{Acobe} that $g_s \approx 0.27$.

Let us next estimate the  possible error resulting from the assumption that the complex structure moduli are perturbed from their minimum of the flux induced potential during inflation. In a large complex structure point generic statements independent of the number of complex structure moduli and the geometry is possible \cite{Brodie:2015kza,Marsh:2015zoa}.
The superpotential to good approximation is given by
\ba\label{complexstruture}
& W \sim  \;\; \cK(\tilde Y_3)_{\alpha \beta \gamma} \, z^\alpha z^\beta z ^\gamma  \;\;, \;\; \nonumber  \\ K=& - \text{log} \big( \tfrac{1}{6}\, \cK(\tilde Y_3)_{\alpha \beta \gamma}  \, \text{Re} z^\alpha \,  \text{Re} z^\beta \,  \text{Re} z ^\gamma  \big)\;\; ,
\ea
where $\tilde Y_3$ is the mirror dual Calabi-Yau threefold to $Y_3$ and thus $\al  =1,\dots, h^{2,1}(Y_3)$. The masses of the complex structure moduli in a non super symmetric vacuum were shown to be   $\mathcal{O} (m_{3/2})$ \cite{Marsh:2015zoa}. Thus we assert that the complex structure modulus perturbed from its minimum  is to admit a mass of the same order as the gravitino mass. With the Hubble scale $H = \sqrt{V_F /3}$ during inflation  and \eqref{scalingMasses} one computes 
\beq\label{m32toHubble}
\frac{H}{m_{3/2}} \, \approx \,  0.039 \, g_s^{\tfrac{1}{2}}  \frac{ |\gamma \cZ|^{\tfrac{3}{2}}}{|\hat \xi|} \, \sim g_s^{2} \;\;.
\eeq
One infers from \eqref{m32toHubble} that for weak string coupling and  moderate values $ \cZ /\chi$ of the geometric background $ H \,/\, m_{3/2} \ll 1$.
Furthermore, one estimates from  \eqref{complexstruture} for the   displacement of $\phi_z $  the complex structure field with canonical kinetic term that $\delta \phi_z \, / \, \langle \phi_z \rangle  \ll 1$. We conclude that the perturbation of the massive complex structure field from the minimum  is small, i.e.\,\,the single field inflation approximation is applicable.

Note that  in the generic moduli case  one encounters an analogous scenario. In the regime \eqref{boundnflation}  a single effective direction in moduli space is not stabilized which  plays the role of the inflaton. We again assume that the complex structure moduli are  perturbed from their vacuum expatiation value in \eqref{W0ds} during inflation such that the lower bound \eqref{boundnflation}  is appropriately obtained. When inflation ends the fluxes acquire their vacuum expectation value and all the K\"ahler moduli are thus stabilized  in the resulting minimum  $\Omega^2 = 0$ for Minkowski and $\Omega^2 \approx 0$ for de Sitter, as discussed in section \ref{sec-Generic}. A more concise analysis of the generic moduli case is desirable.

Let us conclude with some remarks. Firstly, at the end of inflation the modulus is to be stabilized in the potential. The success of this endeavor depends on details of the inflation process.  In particular, on the value of $\phi_{end}$ and the  finite transmission amplitude towards the runaway direction  $\cV \to \infty$, which is expected to be maximal right after inflation ends. Such an analysis  might reduce the obtained region of the tensor-to-scalar ratio.
Secondly, all the  scenarios in this work are crucially  bound to the existence of the $\alpha'$-corrections. However, their ultimate faith is still under investigation.
Let us end on a positive note. It would be of great interest to realize the present scenarios in explicit models, i.e.\,\,geometric backgrounds with suitable topological quantities.

\vspace{1 cm}
\acknowledgements

Many thanks to  Ralph Blumenhagen and Mashito Yamazaki for constructive  and encouraging comments on the draft. 
 In particular, I want to take the opportunity to express my gratitude to the Kavli IPMU for providing  ideal circumstances for inspired research. This work was supported by the WPI program of Japan. \vspace{1 cm}
 
\appendix
\section{Details}\label{Computation}\vspace{-0.3cm}

The intersection numbers of  $Y_3$ are given by
\ba\label{intersectionNumbers}\vspace{-0.3cm}
\cK_{ijk}  &= \int_{B_3} \om_i  \we   \om_j \we   \om_k  \  \, , \;\;\;
\cK_{ij }   = \cK_{ijk}v^k \,,  \nonumber \\
 \cK_{i }   & = \fr12 \cK_{ijk} v^i v^j   \;\;, \;\;\;
\cV   = \fr1{3!} \cK_{ijk}v^i v^j v^k \, .  
\ea 
where $\{\om_i\}$ are the $h_+^{1,1}(Y_3)$ harmonic $(1,1)$-forms.

\nocite{*}
\bibliographystyle{utcaps}
\newpage
\bibliography{references}

\end{document}